\documentclass[prd,twocolumn,showpacs,nofootinbib]{revtex4}
\usepackage{times}
\usepackage{natbib}
\usepackage{epsfig}

\newcommand{\bdv}[1]{{\bf #1}}

\newcommand{\beeq}{\begin{equation}}
\newcommand{\eneq}{\end{equation}}
\newcommand{\bear}{\begin{eqnarray}}
\newcommand{\enar}{\end{eqnarray}}
\newcommand{\up}[1]{{\rm #1}}
\newcommand{\OM}{\Omega_m}
\newcommand{\OB}{\Omega_b}

\newcommand{\rms}{\sigma_8}
\newcommand{\mpc}{{\rm Mpc}}
\newcommand{\mpci}{{\mpc^{-1}}}

\newcommand{\gbar}{\bar g}
\newcommand{\HT}{E^\up{T}}
\newcommand{\dy}{d\chi}
\newcommand{\dobs}{\delta_\up{obs}}
\newcommand{\nobs}{n_g}
\newcommand{\Vang}{\bdv{\hat n}}
\newcommand{\Sang}{\bdv{\hat s}}
\newcommand{\Kang}{\bdv{\hat k}}
\newcommand{\dz}{\delta z}
\newcommand{\ang}{\hat\nabla}
\newcommand{\PP}{P}
\newcommand{\dL}{\mathcal{D}_L}
\newcommand{\dLf}{D_L}
\newcommand{\ddL}{\delta\mathcal{D}_L}
\renewcommand{\aa}{\mathcal{A}}
\newcommand{\ff}{\mathcal{F}}
\newcommand{\vt}{\vartheta}
\newcommand{\np}{n_p}
\newcommand{\II}{\mathcal{T}_l}
\newcommand{\TT}{\Theta}
\newcommand{\GD}{\mathcal{I}}
\newcommand{\dobst}{\delta_\up{obs}^\up{2D}}

\begin{document}

\title{New perspective on galaxy clustering as a cosmological probe:
general relativistic effects}

\author{Jaiyul Yoo$^1$}
\altaffiliation{jyoo@cfa.harvard.edu} 
\author{A. Liam Fitzpatrick$^2$}
\author{Matias Zaldarriaga$^{1,3,4}$}
\affiliation{$^1$Harvard-Smithsonian Center for Astrophysics, Harvard 
University, 60 Garden Street, Cambridge, MA 02138}
\affiliation{$^2$Department of Physics, Boston University, 
590 Commonwealth Avenue, Boston, MA 02215}
\affiliation{$^3$Jefferson Physical Laboratory, Harvard University, 
17 Oxford Street, Cambridge, MA 02138}
\affiliation{$^4$School of Natural Sciences,
Institute for Advanced Study, Einstein Drive, Princeton, NJ 08540}

\begin{abstract}
We present a general relativistic description of galaxy clustering
in a FLRW universe.
The observed redshift and position of galaxies are affected by the matter
fluctuations and the gravity waves
between the source galaxies and the observer, and 
the volume element constructed by using
the observables differs from the physical volume
occupied by the observed galaxies. Therefore, the observed galaxy fluctuation 
field
contains additional contributions arising from the distortion in observable
quantities and these include tensor contributions as well as numerous
scalar contributions. We generalize the linear bias approximation to relate the
observed galaxy fluctuation field to the underlying matter distribution
in a gauge-invariant way. Our full formalism is essential 
for the consistency of theoretical predictions.
As our first application, we compute the angular auto correlation of
large-scale structure and its cross correlation with CMB temperature 
anisotropies. We comment on the possibility of detecting primordial
gravity waves using galaxy clustering and discuss further applications of our
formalism.
\end{abstract}


\maketitle

\section{Introduction}
\label{sec:intro}
Galaxies are known to trace the underlying matter distribution on large
scales and galaxy redshift measurements can therefore provide
crucial information about the time evolution
of large-scale structure of the universe. Over the past decade rapid progress
has been made in this field, following the advent of large galaxy redshift
surveys such as the Sloan Digital Sky Survey (SDSS; \cite{YOADET00}) and the
Two Degree Field Galaxy Redshift Survey (2dFGRS; \cite{CODAET01}),
and much higher precision measurements with
large survey area have opened a new horizon for the role of galaxy clustering
as a cosmological probe (see, e.g., \cite{EIANET01,TESTET06}).

However, a critical question naturally arises: is the Newtonian description
of galaxy clustering  sufficiently accurate
on large scales, close to the horizon scale at high redshift?
General relativity provides
a natural framework for cosmology, and the general relativistic
description is essential to understand the formation of CMB anisotropies, 
since the horizon size at the recombination epoch is of order three degrees 
on the sky (see, e.g., \cite{HUSEET98}).
Therefore, one would naturally expect that a similar relativistic treatment
of galaxy clustering needs to be considered when using galaxy clustering
as a cosmological probe.

Galaxies are measured by observing photons emitted from them and the photon
path is distorted by the matter fluctuations and the gravity waves
 between the source
galaxy and the observer. The volume element constructed using the observed
redshift and observed
angle is different from the real physical volume that the
observed galaxies occupy, and the observed flux and redshift of the source
galaxies are also different from their intrinsic properties.
Therefore, the observed galaxy number density is affected by the
same perturbations given the total number of observed galaxies,
and it contains additional contributions from the distortion in the observable
quantities, compared to the standard description
that galaxies simply trace the underlying matter distribution
$\delta_g=b~\delta_m$. 

Furthermore, perturbations are gauge-dependent quantities and hence 
they are not directly observable. For example, 
the matter fluctuation~$\delta_m$ computed in 
the synchronous gauge is different from the matter fluctuation~$\delta_m$
computed in the conformal Newtonian gauge, which diverges on large scales.
The observable quantities such as observed galaxy clustering should be
independent of a choice of the gauge condition, and this implies that
the standard description is incomplete. Related to this problem is
the linear bias approximation. A scale-independent galaxy bias factor~$b$ 
assumed in one gauge
appears as a scale-dependent galaxy bias factor~$b(k)$ in another gauge.
Galaxy formation is a local process and its relation to the underlying
matter density should be well defined and gauge-invariant.

These issues are naturally resolved when we construct theoretical predictions
in terms of observable quantities. 
Here we provide a fully general relativistic description of galaxy clustering
in a general Friedmann-Lema\^\i tre-Robertson-Walker 
(FLRW) universe, and as our first application we compute the cross correlation
of CMB temperature anisotropies with large-scale structure.
Tracers of large-scale structure contain additional contributions from
relativistic effects and these effects on the cross correlation
progressively become significant at low angular multipoles at high redshift,
since the relativistic effects are significant at the horizon scale
and the horizon size decreases with redshift.
For a photometric quasar sample from the SDSS, we find that
the predicted signals are larger than the standard method would predict
at low angular multipoles and its deviation is larger than the estimated
cosmic variance limit.

The organization of this paper is as follows. In Sec.~\ref{sec:geo}
we describe our notation for a general FLRW metric and solve the geodesic
equation for photons. In Sec.~\ref{sec:lum} we discuss the fluctuation in
luminosity distance that affects the observed flux of source galaxies,
and we present our main results on the general relativistic description 
of galaxy clustering in Sec.~\ref{sec:ngal}. In Sec.~\ref{sec:cross}
we compute the angular correlation of large-scale structure and its 
cross correlation with CMB anisotropies with the main emphasis on the 
systematic errors. Finally, we discuss the implication of our new results
and conclude with a discussion of further applications in
Sec.~\ref{sec:discussion}.

\section{Geodesic equation}
\label{sec:geo}
We present our notation for the background metric in an inhomogeneous
universe and solve the geodesic equation for photons emitted from galaxies
to derive the relation between the source galaxies and the observer.

\subsection{FLRW Metric}
\label{ssec:metric}
We assume that the background universe is well
described by the Friedmann-Lema\^\i tre-Robertson-Walker (FLRW)
metric with a constant spatial curvature,
\beeq
ds^2=g_{ab}~dx^adx^b=-dt^2+a^2(t)~\gbar_{\alpha\beta}~dx^\alpha dx^\beta~,
\eneq
where $a(t)$ is the scale factor and
$\gbar_{\alpha\beta}$ is the metric tensor for a three-space.
The conformal time $\tau$ is defined as $a~d\tau=dt$ with the speed of
light $c\equiv1$, and it is related to the comoving line-of-sight distance,
\beeq
r(\tau)=a(\tau_0)(\tau_0-\tau)=a(\tau_0)\int_t^{t_0}{dt\over a(t)}=
\int_0^z{dz\over H(z)}~,
\eneq
where $H=\mathcal{H}/a=\dot a/a^2$ 
is the Hubble parameter and the dot denotes the
derivative with respect to the conformal time. 
The subscript~$0$ represents that the quantities are computed at origin in
a homogeneous universe. In a flat universe,
the comoving line-of-sight distance is coincident with the comoving
angular diameter distance. From now on we set $a(\tau_0)\equiv1$. 

The metric tensor can be expanded to represent its perturbations
for the spacetime geometry and to describe the departure from the homogeneity
and isotropy,
\bear
\label{eq:metric}
ds^2&=&-a^2\left(1+2A\right)d\tau^2-2~a^2B_\alpha ~d\tau ~dx^\alpha \\
&+&a^2\left[(1+2D)\gbar_{\alpha\beta}+2E_{\alpha\beta}\right]dx^\alpha dx^\beta~.
\nonumber
\enar
We can further decompose the perturbation variables depending on their
spatial transformation properties as $B_\alpha=B~Q_\alpha$ and
$E_{\alpha\beta}=E~Q_{\alpha\beta}+\HT_{\alpha\beta}$,
where $\HT_{\alpha\beta}$ is the divergenceless tensor.
We adopted the convention \cite{BARDE80} 
for the eigenmode~$Q_\alpha$ and~$Q_{\alpha\beta}$ of the Helmholtz equations
and assumed there is no vector mode.
Throughout the paper we use Greek indices to represent 
the 3D spatial components, running from~1 to~3, while Latin indices are used 
to represent the 4D spacetime components with~0 being the conformal
time component.

Here we will
work with the general representation of the metric without fixing gauge 
conditions (see, e.g., \cite{BARDE80,KOSA84,HWNO01}),
but it often proves convenient to understand our general formulas in
conjunction with other gauges such as the conformal Newtonian gauge
and the synchronous gauge. The metric in the conformal Newtonian gauge 
(see, e.g., \cite{MUFEBR92}) is
\beeq
\label{eq:newton}
ds^2=-a^2\left(1+2\psi\right)d\tau^2 
+a^2\left[(1+2\phi)\gbar_{\alpha\beta}
+2\HT_{\alpha\beta}\right]dx^\alpha dx^\beta~, 
\eneq
and the metric in the synchronous gauge (see, e.g., \cite{MABE95}) is
\beeq
ds^2=-a^2d\tau^2+a^2\left[\gbar_{\alpha\beta}+h_{\alpha\beta}\right]
~dx^\alpha dx^\beta~.
\label{eq:metric-sync}
\eneq
Throughout the paper, we adopt as our fiducial model a flat $\Lambda$CDM 
universe with the matter density $\OM=0.24$ ($\OM h^2=0.128$), 
the baryon density $\OB=0.042$ ($\OB h^2=0.0224$), the Hubble constant $h=0.73$,
the spectral index $n_s=0.954$, the optical depth to
the last scattering surface $s=0.09$, and the primordial
curvature perturbation amplitude $\Delta_\phi^2=2.38\times10^{-9}$
at $k=0.05~\mpci$ ($\rms=0.81$), consistent with the
recent results (e.g., \citep{TESTET06,KODUET08}).
We use the Boltzmann code CMBFast \citep{SEZA96} to obtain the transfer 
functions of perturbation variables.

\subsection{Temporal component: Sachs-Wolfe effect}
The photon geodesic
$x^a(\lambda)$ can be parametrized by an affine parameter $\lambda$, and 
its propagation direction is then $k^a=dx^a/d\lambda$, subject to
the null equation ($ds^2=k^ak_a=0$). We choose the normalization of 
the affine parameter, such that the time component of the null vector
represents the photon frequency $\bar\nu$, measured by an observer
in a homogeneous universe \cite{YOO09}.
The null vector is therefore
\beeq
k^0={\bar\nu\over a}~(1+\delta\nu),~~~
k^\alpha=-~{\bar\nu\over a}~(e^\alpha+\delta e^\alpha),
\label{eq:null}
\eneq
where the unit vector $e^\alpha$ is the photon propagation direction seen
from the 
observer. The spatial component of the null vector is obtained by the
null condition and we expanded the null vector to the first order in 
perturbations to represent its dimensionless temporal and spatial
perturbations~$\delta\nu$ and $\delta e^\alpha$.

To the zeroth order in perturbations, the photon frequency is redshifted as
$\bar\nu\propto1/a$ in an expanding universe, and the geodesic path is
described by 
$d/\dy\equiv(a/\bar\nu)(d/d\lambda)=\partial_\tau-e^\alpha\partial_\alpha=-d/dr$. 
Equivalently the affine parameter~$\chi$ describes the same geodesic path
$x^a(\chi)$, but in a conformally transformed metric
$\tilde g_{ab}=(\bar\nu/a)~g_{ab}$ (see, e.g., \cite{WALD84} for conformal
transformation). We will put tilde to represent quantities
in the conformally transformed metric. 

The temporal component of the null vector can be integrated to 
obtain the relation between~$\tau$ and~$\chi$ as
\beeq
\tau-\tau_o=\chi-\chi_o+\int_{\chi_o}^{\chi}\dy'~\delta\nu(\chi')~,
\label{eq:tauchi}
\eneq
where the subscript~$o$ indicates that the affine parameter is computed at
origin in an inhomogeneous universe.
The perturbations of the null vector are related to the metric perturbations as
\beeq
e^\alpha~\delta e_\alpha=\delta\nu+A-B_\alpha ~e^\alpha-D-E_{\alpha\beta}~e^\alpha
e^\beta~,
\label{eq:nc}
\eneq
by the null equation, and as
\beeq
{d\over\dy}(\delta\nu+2A)=(\dot A-\dot D)-(B_{\alpha|\beta}+
\dot E_{\alpha\beta})~e^\alpha e^\beta~,
\label{eq:zerogeo}
\eneq
by the temporal component of the geodesic equation ($k^0{_{;b}}k^b=0$).
The vertical bar and the semicolon represent
the covariant derivatives with respect to $\gbar_{\alpha\beta}$ and $g_{ab}$,
respectively.

Consider a comoving observer of which the rest frame has vanishing
total three momentum. Its four velocity  is $u^a=[(1-A)/a,~v^\alpha/a]$ and
the observer measures the redshift parameter of a source,
\beeq
1+z_s={(k^a~u_a)_s\over(k^a~u_a)_o} 
=\left({a_o\over a_s}\right)\bigg\{1+
\big[\delta\nu+A+(v_\alpha-B_\alpha)~e^\alpha\big]^s_o\bigg\}~, 
\eneq
with the spacetime of the source indicated by the subscript~$s$ and
the bracket representing a difference of the quantities at two spacetime
points.
Using Eq.~(\ref{eq:zerogeo}), this relation can be further simplified 
\cite{HWNO99} as 
\bear
\label{eq:gSW}
1+z_s&=&\left({a_o\over a_s}\right)\bigg\{
1+\bigg[(v_\alpha-B_\alpha)~e^\alpha-A\bigg]_o^s \\
&-&\int_0^{r_s}dr\left[(\dot A-\dot D)-(B_{\alpha|\beta}+\dot E_{\alpha\beta})
~e^\alpha e^\beta\right]\bigg\}~, \nonumber 
\enar
where $r_s=r(z_s)$ is the comoving line-of-sight distance to the source 
galaxies at~$z_s$ and $v_\alpha e^\alpha$ is the line-of-sight peculiar velocity.
Equation~(\ref{eq:gSW}) in the conformal Newtonian gauge is
known as the Sachs-Wolfe effect \citep{SAWO67}.
The first square bracket represents the redshift-space
distortion by peculiar velocities, frame dragging, and gravitational redshift,
respectively. The first round bracket in the integral also
represents the gravitational redshift,
arising from the net difference in gravitational potential due to its
time evolution for the duration of photon propagation, and this effect is
referred to as the integrated Sachs-Wolfe effect. The last terms in the
integral represent the tidal effect from the frame dragging 
and the integrated Sachs-Wolfe effect
from the time evolution of the primordial gravity waves.

Since the redshift parameter in a homogeneous universe is defined
as $1/a$, we define a quantity $\dz$ that relates the observed
redshift~$z_s$ of the source and the redshift of the source
that would be measured in a homogeneous universe as
$1/a_s\equiv(1+z_s)(1-\dz)$, and note that
$a_o=1+\mathcal{H}_o\delta\tau_o$.
The redshift $1/a_s$ of the source in a homogeneous universe is not directly
measurable and hence $\dz$ is gauge-dependent.
One can easily verify that for a coordinate transformation
$\tau\rightarrow \tau'=\tau+T$, the perturbation in the observed
redshift transforms as $\dz\rightarrow\dz'=\dz+\mathcal{H}T$,
while the observed redshift~$z_s$ is gauge-invariant.\footnote{Particular
attention needs to be paid to the difference between~$z$ and~$1/a$
in conjunction with Eq.~(\ref{eq:gSW}). Throughout the paper the redshift
parameter~$z$ refers to the ``observed'' redshift, which is different
from the gauge-dependent redshift parameter $z_h$ in a homogeneous
and isotropic universe, defined as $1+z_h=1/a$~.}

\subsection{Spatial components: Gravitational lensing effect}
Metric perturbations, sourced by matter fluctuations and gravity waves
along the line-of-sight,
deflect the photon propagation direction emitted from galaxies and
displace their observed position on the sky. This effect, known as the
gravitational lensing effect, is described by the spatial components
of the geodesic equation $(k^\alpha{_{;b}}k^b=0)$ as 
\bear
\label{eq:ageo}
&&{d\over\dy}~(\delta e^\alpha+B^\alpha+2D ~e^\alpha+2E_\beta^\alpha~e^\beta)  \\
&&=\delta e^\beta~ e^\alpha{_{|\beta}}-\delta\nu~\dot e^\alpha+
A^{|\alpha}-B_\beta{^{|\alpha}}e^\beta 
-D^{|\alpha}-E_{\beta\gamma}^{|\alpha}~e^\beta~ e^\gamma~. \nonumber
\enar
Noting that $(d/\dy)~\delta x^\alpha=-\delta e^\alpha$, the spatial
components of the geodesic equation can be integrated and expressed
in spherical coordinates to obtain the angular displacements
\bear
&&\delta\theta=-\int_0^{r_s}dr~\bigg\{{\left[(B^\alpha-B^\alpha_o)
+2(E^{\alpha\beta}-E^{\alpha\beta}_o)e_\beta\right] e^\theta_\alpha
\over r_s} \nonumber \\
&&+\left({r_s-r\over r~r_s}\right){\partial\over\partial\theta}\left(A-D
-B_\alpha ~e^\alpha-E_{\alpha\beta} ~e^\alpha~e^\beta\right)\bigg\}~, \hspace{25pt}
\enar
and 
\bear
&&\delta\phi=-\int_0^{r_s}dr~\bigg\{{\left[(B^\alpha-B^\alpha_o)
+2(E^{\alpha\beta}-E^{\alpha\beta}_o)e_\beta\right]
e^\phi_\alpha\over r_s\sin\theta} \nonumber \\
&&+\left({r_s-r\over r~r_s\sin^2\theta}\right)
{\partial\over\partial\phi}
\left(A-D-B_\alpha ~e^\alpha-E_{\alpha\beta} ~e^\alpha~e^\beta\right)
\bigg\}~.  \hspace{20pt}
\enar
Apart from the frame distortion described by 
$\GD^\alpha\equiv(B^\alpha-B^\alpha_o)+2(E^{\alpha\beta}-E^{\alpha\beta}_o)e_\beta$,
the gravitational lensing displacement depends only on the spatial derivative
of the metric perturbations, i.e., a constant gravitational potential
results in no observable effect.

Since the comoving line-of-sight distance to the source 
in an inhomogeneous universe is $(\tau_0-\tau_s)$ and
the source position~$\tau_s$ is related to
the observed redshift~$z_s$ through $a_s$
in Eq.~(\ref{eq:gSW}), it can be expressed 
in terms of $r(z)$ and $H(z)$ in a homogeneous universe as
\beeq
\label{eq:cotau}
\bar r\equiv\tau_0-\tau_s=r\left[z_s-(1+z_s)~\dz\right] 
=r_s\left(1-{1+z_s\over H_sr_s}~\dz\right)~, 
\eneq
where $H_s=H(z_s)$. Note that we have expanded the argument 
of $r(x)$ in the square bracket 
around the observed redshift~$z_s$ of the 
source. Finally, using the null equation, the radial displacement is then
obtained as
\bear
\label{eq:deltar}
\delta r&=&\chi_o-\chi_s+e_\alpha~\delta x^\alpha-\bar r \\
&=&\delta\tau_o+\int_0^{r_s}dr~\left(A-D-B_\alpha~ e^\alpha+E_{\alpha\beta}~
e^\alpha e^\beta\right)~.\nonumber
\enar

With the full solution of the geodesic equation, the angular position~$\Sang$
of the source galaxies can be obtained by tracing backward the photon path
and expressed in terms
of observed angle $\Vang=(\theta,~\phi~\sin\theta)$ as
$\Sang=[\theta+\delta\theta,~(\phi+\delta\phi)~\sin(\theta+\delta\theta)]$.
Due to the lensing displacement
a unit solid angle $|d^2\Sang|$ in the source plane is distorted to
a unit solid angle $|d^2\Vang|$ in the image plane.
The amplitude of this distortion is described by
the convergence~$\kappa$ as
\beeq
\label{eq:magni}
\left|{d^2\Vang\over d^2\Sang}\right|
=1-{\partial\over\partial\phi}~\delta\phi-
\left(\cot\theta+{\partial\over\partial\theta}\right)
\delta\theta\equiv1+2~\kappa~,
\eneq
and therefore
\bear
\label{eq:kappa}
&&\kappa=\int_0^{r_s}dr~\bigg\{
{\csc\theta~\partial_\phi (e^\phi_\alpha ~\GD^\alpha)
+\partial_\theta (e^\theta_\alpha ~\GD^\alpha)
+\cot\theta ~e^\theta_\alpha ~\GD^\alpha
\over 2~r_s} \nonumber \\
&&+\left({r_s-r\over 2~r~r_s}\right)\ang^2
\left(A-D-B_\alpha ~e^\alpha-E_{\alpha\beta} ~e^\alpha~e^\beta\right)
\bigg\}~, \hspace{25pt}
\enar
where $\ang$ is the differential operator in two dimensional unit sphere.
In the literature Eq.~(\ref{eq:magni})
is often referred to as the gravitational 
lensing magnification~$\mu$. However, the angular position~$\Sang$ of the
source galaxies is not observable; its coordinate value depends on the choice
of gauge condition, while the spacetime of the source position is physical.
Consequently, the convergence~$\kappa$ in Eq.~(\ref{eq:kappa})
is gauge-dependent, whereas magnification should be a gauge-invariant
quantity. In Sec.~\ref{sec:lum}
we provide a correct gauge-invariant expression for magnification~$\mu$.
Note that the gravitational lensing displacements $\delta r$, $\delta\theta$,
and $\delta\phi$ are also gauge-dependent.

The standard Newtonian
expression for the convergence can be obtained with a few
approximations: When the Newtonian potential and curvature are constant 
in time as in an Einstein-de~Sitter universe, we can replace the total
derivative $d/dr$ by the partial derivative~$\partial_r$. Integrating by part
and ignoring the boundary terms yield the standard form 
\cite{JASEWH00,HISE03a} as
\bear 
\kappa&=&\int_0^{r_s}dr~{(r_s-r)~r\over 2~r_s}~
\left[\nabla^2-{1\over r^2}{\partial\over\partial r}
\left(r^2{\partial\over\partial r}\right)\right]
\left(\psi-\phi\right) \nonumber \\
&=&{3H_0^2\over2}~\OM\int_0^{r_s}dr~
{\delta_m\over a}~{(r_s-r)~r\over r_s}~,
\enar
where we have used the Newtonian Poisson equation
$k^2\phi\simeq-k^2\psi\simeq4\pi G~a^2\delta\rho$.
Deep inside the horizon where the Newtonian approximation is accurate,
there is no gauge ambiguity and the gravitational lensing magnification is
$\mu\simeq|d^2\Vang/d^2\Sang|=1+2~\kappa$~.

\section{Observed Luminosity distance}
\label{sec:lum}
The observed position and the redshift of source galaxies are affected by
the matter fluctuations and the gravity waves
between the source galaxies and the observer,
and this relation is described by the geodesic equation in Sec.~\ref{sec:geo}.
The observed flux of the source galaxies
is also affected by the same fluctuations
and this relation is described by the fluctuations in the luminosity distance.
Here we derive the observed luminosity distance in an inhomogeneous universe
(see \cite{SASAK87,BODUGA06} for earlier derivations).

Consider a source with intrinsic luminosity $L$ and proper radius
$\Delta R_s$. The flux $\ff_o$ and redshift~$z_s$
of the source are measured at origin and
the observed luminosity distance is defined as
\beeq
\dL(z_s)\equiv\sqrt{L\over4\pi \ff_o}=\sqrt{\ff_s\over \ff_o}~\Delta R_s=
{\aa_s\nu_s\over \aa_o\nu_o}~\Delta R_s~,
\label{eq:dl}
\eneq
where $\aa$ is the scalar amplitude of the four potential of the photons
and we have used $\ff\propto\aa^2\nu^2$. When the wavelength of the photons
is shorter than the curvature scale, the propagation of light rays can
be locally described by Maxwell's equations, and the governing equations
are known as 
the geometric optics in curved spacetime (see, e.g., \cite{SACHS61,MTW}).

The optical scalar equations are the propagation
equations of the scalar amplitude 
\beeq
{d\over \dy}\left(\aa~a\right)+{1\over2}~\aa a\vt=0~,
\label{eq:aa}
\eneq
and the expansion of the wavevector $\vt=\tilde k^a{_{;a}}$ 
\beeq
{d\over\dy}~\vt+{1\over2}~\vt^2=-\tilde R_{ab}\tilde k^a\tilde k^b~,
\label{eq:vt}
\eneq
where $\tilde R_{ab}$ is the Ricci tensor in the conformally transformed
metric $\tilde g_{ab}=(\bar\nu/a)~g_{ab}$~.
To the zeroth order in perturbations Eq.~(\ref{eq:vt}) has no source term
in a flat universe and it can be integrated to obtain
the expansion of the wavevector 
 $\vt=2/(\chi-\chi_s-\Delta\chi_s)$, where $\Delta\chi_s$ is related to
 the size of the source. Since the proper radius of the
source is $\Delta R_s=|dt|$ in a local Lorentz frame, it can be expressed
in terms of the affine parameter~$\chi$ by considering the photon frequency
at the source as
\beeq
\label{eq:dt}
-\nu_s=(k^au_a)_s=-{\bar\nu_s\over a_s}~{dt\over \dy}~.
\eneq
Note that $dt$ in Eq.~(\ref{eq:dt}) is defined in the local Lorentz frame
of the source.
Solving Eq.~(\ref{eq:aa}) for $\aa a$ and using 
$\Delta R_s=a_s\nu_s|\Delta\chi_s|/\bar\nu_s$ yields
the observed luminosity distance as
\bear
\label{eq:dl2}
\dL(z_s)&=&(1+z_s)\Delta R_s~{a_o\over a_s}~\exp\left[-\int_o^s
\dy~{\vt\over2}~\right] \\
&=&a_o~(1+z_s)~(\chi_o-\chi_s)~{\nu_s\over\bar\nu_s}~\left(1-\int_o^s
\dy~{\delta\vt\over2}~\right)~, \nonumber 
\enar
in the limit $\Delta\chi_s\rightarrow0$, and
 $\delta\vt$ is the first order perturbation of the expansion
of the wavevector that can be obtained by expanding Eq.~(\ref{eq:vt}).

Now to solve for $\delta\vartheta$ we integrate Eq.~(\ref{eq:vt})
along the zeroth order solution $\vt$,
\beeq
\label{eq:lenskappa}
\int_o^s\dy~{\delta\vt\over2}=
\int_0^{r_s}dr~{(r_s-r)~r\over2~ r_s}~\delta(\tilde R_{ab}\tilde k^a
\tilde k^b)~,
\eneq
with the source term in the integral 
\bear
\delta(\tilde R_{ab}\tilde k^a\tilde k^b)&=&
-k^2\left[A-\left(D+{E\over3}\right)+
\left({\dot B\over k}-{\ddot E\over k^2}\right)\right] \nonumber \\
&-&2\left(\ddot D+{\ddot E\over3}\right) 
+4\left(\dot D+{\dot E\over3}\right)_{|\alpha}e^\alpha \nonumber \\
&-&\left[A+\left(D+{E\over3}\right)+
\left({\dot B\over k}-{\ddot E\over k^2}\right)\right]_{|\alpha\beta}
e^\alpha e^\beta \nonumber \\
&+&\left(\ddot E^{T}_{\alpha\beta}+k^2E^{T}_{\alpha\beta}\right)e^\alpha e^\beta~.
\enar
Noting that the luminosity distance in a homogeneous universe is 
$\dLf(z)=(1+z)~r(z)$ and the 
comoving line-of-sight distance~$\bar r$ of the source
is related to the affine parameter via
Eqs.~(\ref{eq:tauchi}) and~(\ref{eq:cotau}),
the observed luminosity distance $\dL(z_s)$
can be written as \cite{SASAK87}
\bear
\label{eq:dLfull}
&&{\dL(z_s)\over\dLf(z_s)}=1+(v_\alpha-B_\alpha)_se^\alpha
-A_s-{1+z_s\over H_s~r_s}~\dz \nonumber \\
&&+2\int_0^{r_s}dr{A\over r_s} 
-\int_0^{r_s}dr{r\over r_s}\left[(\dot A-\dot D)-
(B_{\alpha|\beta}+\dot E_{\alpha\beta})e^\alpha e^\beta\right] \nonumber \\
&&-\int_0^{r_s}dr~{(r_s-r)~r\over2~ r_s}~\delta(\tilde R_{ab}\tilde k^a
\tilde k^b)
+\left(\mathcal{H}_o+{1\over r_s}\right)\delta\tau_o~.
\hspace{20pt}
\enar

With the full expression for luminosity distance, the magnification of
a source at observed redshift~$z$ is defined
as the ratio of the observed flux~$\ff_o$ to the flux of the source
that would be measured in a homogeneous universe:
\beeq
\mu=\ff_o\left({L\over4\pi\dLf^2}\right)^{-1}=
\left({\dLf\over\dL}\right)^2=1-2~\ddL~.
\eneq
We have defined the perturbations in Eq.~(\ref{eq:dLfull}) as
$\dL(z)\equiv\dLf(z)(1+\ddL)$, and note that written in terms of observable
variables $\ddL$ is gauge-invariant and both $\dL(z)$ and $\dLf(z)$
are evaluated at the observed redshift~$z$.
In the Newtonian limit,
Eq.~(\ref{eq:lenskappa}) becomes the convergence~$\kappa$ and it is the
dominant factor for $\ddL$. Therefore, we recover the Newtonian expressions
 $\ddL\simeq-\kappa$ and $\mu\simeq1+2~\kappa$~.

\section{Observed galaxy fluctuation field}
\label{sec:ngal}
Drawing on the formalism developed in Sec.~\ref{sec:geo} and~\ref{sec:lum}
we present the expression for the observed galaxy fluctuation field~$\dobs$,
accounting for all the relativistic effects to the linear order. Our formalism
is crucial for the theoretical consistency and the gauge-invariance of the
predictions using galaxy clustering as a cosmological probe.
To construct the observed galaxy overdensity field we start by considering 
a gauge-invariant quantity, the total number $N_\up{tot}$ of observed
galaxies. The total number of observed galaxies in a small volume
described by observed redshift~$z$ and observed angle~$\Vang$
can be formulated in terms of a covariant volume integration 
in a four-dimensional spacetime manifold \cite{WEINB72},
and it is related to the photon geodesic~$x^a(\chi)$ via 
\beeq
N_\up{tot}=\int\sqrt{-g}~\np~\varepsilon_{abcd}~u^d~
{\partial x^a\over\partial z}{\partial x^b\over\partial\theta}
{\partial x^c\over\partial\phi}~dz~ d\theta~ d\phi~, 
\label{eq:ngal}
\eneq
where $\np$ is the physical number density of the source galaxies, 
the metric determinant is $\sqrt{-g}=a^4~(1+A+3D)$, and
$\varepsilon_{abcd}=\varepsilon_{[abcd]}$ is the Levi-Civita symbol.

To the linear order in perturbations the photon geodesic is a straight
path with small distortion and, with the geodesic path
$x^a(\chi)$ in Sec.~\ref{sec:geo} we obtain
\bear
&&N_\up{tot}=\int \np~
{r^2\sin\theta \over (1+z)^3H}~dz ~d\theta ~d\phi
~\bigg[1+3D+v^\alpha e_\alpha+2~{\delta r\over r} \nonumber  \\
&&+H{\partial\over\partial z}~\delta r
+\left(\cot\theta+{\partial\over\partial\theta}\right)
\delta\theta+{\partial\over\partial\phi}~ \delta\phi 
+{\bar r^2\over r^2}~H~{\partial\bar r\over\partial z}
\bigg]. 
\enar
The observed galaxy number density~$\nobs$ is then defined in relation 
to the total number of observed galaxies and the observed volume element as
\beeq
N_\up{tot}\equiv\int~\nobs~{r^2\sin\theta\over(1+z)^3H}~
~dz~ d\theta~ d\phi~,
\label{eq:nobs}
\eneq
and therefore  the observed galaxy number density is 
\bear
\label{eq:nnn}
\nobs=\np~\bigg[1&+&A+2D+(v^\alpha-B^\alpha)e_\alpha+E_{\alpha\beta}e^\alpha 
e^\beta\nonumber  \\
&-&(1+z){\partial\over\partial z}~\dz-2~{1+z\over Hr}~\dz-\dz\nonumber \\
&-&2~\kappa+
{1+z\over H}{dH\over dz}~\dz
+2~{\delta r\over r}\bigg]~.
\enar
Given the total number of observed galaxies,
the observed galaxy number density is affected by the matter fluctuations
and the gravity waves,
since the volume element is constructed in terms of observed redshift
and observed angle. 
Equation~(\ref{eq:ngal}) automatically takes into account the full effects
of the volume distortion described by the photon geodesic equation in 
Sec.~\ref{sec:geo}. 

However, additional distortions arise due to the intrinsic luminosity
function $d\np/dL$ of the source galaxies. As described in Sec.~\ref{sec:lum}
the observed flux of the source galaxies
is affected by the matter fluctuations and the gravity waves between the
source galaxies and the observer, 
 and magnification of the source galaxy
flux changes the observed galaxy number density.
Given an observational threshold $\ff_\up{thr}$ in flux at origin,
the physical number density $\np$ in the above equations
should be modified as
\beeq
\np\rightarrow
\int_{\ff_\up{thr}}^\infty d\ff_o~{dL\over d\ff_o}~{d\np\over dL}
=\np\left[L_\up{thr}(1+2~\ddL)\right]~,
\eneq
where $\np(L)$ is the cumulative (physical) number density of 
the source galaxies brighter than~$L$ and 
$L_\up{thr}=4\pi\dLf^2(z)\ff_\up{thr}$ is the inferred luminosity threshold
for the source galaxy sample.
For a galaxy sample with $d\np/dL\propto L^{-s}$, the cumulative
number density can be expanded as $\np(L_\up{thr})(1-5~p~\ddL)$,
and $p=0.4~(s-1)$ is the slope of the luminosity function in magnitude.

Furthermore, since we observe galaxies rather than the underlying matter 
distribution, we need to relate the physical number density $\np$ of the
source galaxies to the matter density~$\rho_m$.
In the simplest model of galaxy formation, the galaxy number density is simply
proportional to the underlying matter density $\rho_m$, when $\rho_m$ is
above some threshold $\rho_t$ dictated by complicated but local process
involving atomic physics. The matter density at the source galaxy position is 
related to the mean matter density at the observed redshift~$z$ 
as\footnote{The observed redshift~$z$ is related to the expansion 
parameter~$a$ of the source galaxy as $1+z=(1+\dz)/a$~.}
\beeq
\rho_m(x^a)={\bar\rho_m(\tau_0)\over a^3}\left(1+\delta_m\right) 
=\bar\rho_m(z)\left[1+\delta_m-3~\dz\right]~, 
\eneq
and the mean matter density at the observed redshift is
$\bar\rho_m(z)=(3H_0^2/8\pi G)\OM(1+z)^3$.
The combination $(\delta_m-3~\dz)$ is gauge-invariant and is
proportional to the matter density at the source galaxy position.
Within the linear bias approximation, the long wavelength
fluctuations of the matter density $\rho_m$ at a given point
effectively lower the threshold for galaxy formation and 
the galaxy number density can be written as 
\beeq
\np=\bar \np(z)~\left[1+b~(\delta_m-3~\dz)\right]~,
\label{eq:npp}
\eneq
and~$b$ is a scale-independent 
linear bias factor.\footnote{More general ansatz for Eq.~(\ref{eq:npp})
can be obtained by generalizing the earlier approach \cite{KAISE84,POWI84}
as
$\np=\bar\np(z)~\exp\left[b_L\int \sqrt{-g}~d^4y~(\delta_m-3~\dz)(y)~
\mathcal{W}(x-y)\right]~,$ where $\mathcal{W}$ is a local filter function
that cuts off small scale fluctuations, and the Lagrangian bias~$b_L$ is
related to the bias in Eulerian space as $b=b_L+1$.}
Equation~(\ref{eq:npp}) can be contrasted
with the gauge-dependent relation $\delta_g=b~\delta_m$~.

\begin{figure}
\centerline{\psfig{file=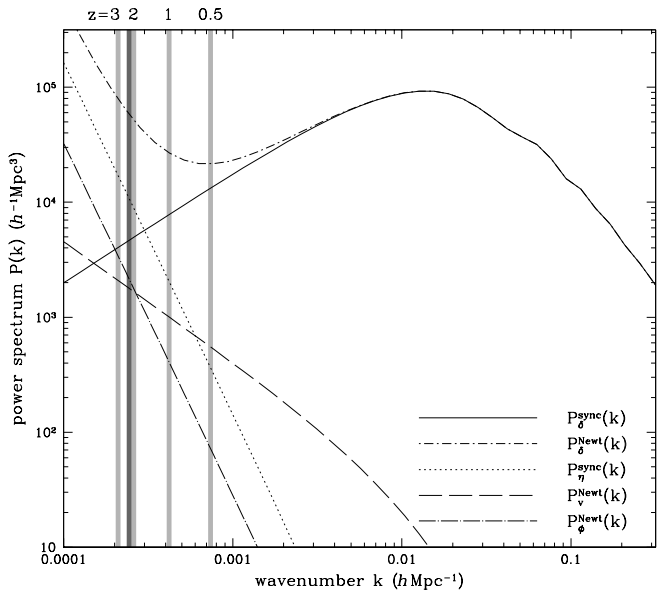, width=3.4in}}
\caption{Power spectra of perturbation variables
computed at $z=0$ in the conformal Newtonian and the synchronous gauges.
Vertical lines show the comoving line-of-sight distance
($k=1/r(z)$; {\it light gray}) at each redshift indicated in the legend and
the horizon scale ($k=H_0$; {\it dark gray}) today. Near the horizon scale,
even power spectra of matter fluctuations in two gauges differ dramatically,
showing that gauge effects are substantial and it is nontrivial to
relate perturbation variables to observable quantities. Two distinct choices
of gauge conditions cannot be used simultaneously
around the horizon scale (e.g., Newtonian
gauge equations with synchronous gauge transfer function
outputs from CMBFast or CAMB).}
\label{fig:pow}
\end{figure}

Finally, putting all the ingredients together the observed galaxy fluctuation 
field can be written as
\bear
\label{eq:dobs}
\dobs&=&b~(\delta_m-3~\dz)+
A+2D+(v^\alpha-B^\alpha)e_\alpha+E_{\alpha\beta}e^\alpha 
e^\beta\nonumber  \\
&-&(1+z){\partial\over\partial z}~\dz-2~{1+z\over Hr}~\dz-\dz
-5p~\ddL-2~\kappa
\nonumber \\
&+&{1+z\over H}{dH\over dz}~\dz+2~{\delta r\over r}~,
\enar
where $\dz$, $\delta r$, $\kappa$, and $\ddL$ are
given in Eqs.~(\ref{eq:gSW}), (\ref{eq:deltar}),~(\ref{eq:kappa}), 
and~(\ref{eq:dLfull}), respectively.
This equation is the main result of our paper.
Constructed from the gauge-invariant expressions and expressed in terms of
observables, this result is  gauge-invariant. Note that
in addition to the scalar contributions,
Eq.~(\ref{eq:dobs}) includes tensor contributions from the primordial
gravity waves, mainly from the integrated Sachs-Wolfe effect in~$\dz$.

One remaining ambiguity in computing~$\dobs$ is the time lapse $\delta\tau_o$
at origin, representing the departure from 
$\tau_0$ in a homogeneous universe. However, this quantity
is independent of the position and angle of the source galaxies;
In practice the mean number density 
$\bar n_p(z)$ of the observed 
galaxies is obtained by averaging $n_g$ over observed angle~$\Vang$
at a fixed observed redshift~$z$
and $\delta\tau_o$ is absorbed in the monopole set equal $\bar n_p(z)$.
In the Newtonian limit the dominant
contribution in $\dz$ is the peculiar velocity~$V$
and Eq.~(\ref{eq:dobs})
reduces to the standard relation for redshift-space distortions \cite{KAISE87}
and magnification bias \cite{NARAY89} as
\beeq
\label{eq:std}
\delta_\up{std}=
b~\delta_m+(5p-2)~\kappa-{1+z\over H}~{\partial V\over\partial r}~.
\eneq

Figure~\ref{fig:pow} illustrates the theoretical inconsistency in 
the standard method by showing the power spectra of perturbation variables
computed at $z=0$ in the conformal Newtonian and the synchronous gauges.
The power spectra of matter fluctuations
in two gauges ({\it solid}; synchronous,
{\it short dot-dashed}; conformal Newtonian) noticeably deviate from each
other well before they reach the horizon scale ({\it dark gray}),
reflecting that theoretical predictions in the standard method depend on
the choice of gauge conditions. In particular, as we observe higher redshift,
larger comoving scales ({\it light gray}) are accessible and the 
horizon scale is smaller, and therefore the systematic errors in the
standard methods start to become significant on progressively smaller scales.
The infrared divergence
shown as the dot-dashed line on large scales is an artifact in the conformal
Newtonian gauge, while the matter fluctuation~$\delta_m$
({\it solid}) in the synchronous gauge is also gauge-dependent.
Theoretical quantities plotted in Fig.~\ref{fig:pow} are not directly 
observable.

\section{Cross correlation of CMB anisotropies with large-scale structure}
\label{sec:cross}
As the first application of our formalism, we compute the angular correlation
of large-scale structure and its cross correlation with CMB anisotropies.
In the standard approach, the observed galaxy fluctuation field is written 
in the Newtonian limit, and neglecting the additional 
contributions to the observed galaxy fluctuation field results in systematic 
errors in the theoretical predictions. We first introduce the formalism
for computing the angular correlations in Sec.~\ref{ssec:obs}, and 
present the angular auto and cross correlations with the main emphasis on
the systematic errors in Sec.~\ref{ssec:ang}.

\subsection{Observed angular fluctuation field}
\label{ssec:obs}
The observed angular fluctuation field can be obtained by integrating
Eq.~(\ref{eq:dobs}) along the line-of-sight as
\beeq
\dobst(\Vang)=\int dz\PP(z)~\dobs(z,\Vang)~, 
\eneq
with the normalized selection function $\PP(z)$ of the galaxy sample.
The selection function $\PP(z)$ can be obtained by averaging the observed
galaxy number density $n_g$ at each observed redshift slice.
Since $\dobs$ in Eq.~(\ref{eq:dobs}) is a linear combination of perturbation 
variables $T_i$ with different weight function $W_i(r,\Vang,\Kang)$,
it proves convenient to further decompose their functional dependence by
\beeq
\dobs(z,\Vang)=\sum_i\int_0^{r_s}dr\int{d^3\bdv{k}\over(2\pi)^3}~
W_i(r,\Vang,\Kang)~T_i(\bdv{k},r)~e^{i\bdv{k}\cdot\bdv{x}}~,
\eneq
and $\bdv{x}=(r,\Vang)$ in the spherical coordinate.
The angular fluctuation field is often expanded
as a function of spherical harmonics and the observed angular component is
then
\bear
a_{lm}&=&\int d^2\Vang~\dobst(\Vang)~Y^*_{lm}(\Vang)=\sum_i
\int{d^3\bdv{k}\over(2\pi)^3} \int dz\PP(z) \nonumber \\
&\times&\int_0^{r_s}dr\int d^2\Vang~Y^*_{lm}(\Vang)~W_i(r,\Vang)
~T_i(\bdv{k},r)~e^{i\bdv{k}\cdot\bdv{x}}.
\enar

For most of the perturbation variables such as $\delta_m$, $A$, and $D$ in
Eq.~(\ref{eq:dobs}), the weight function takes the simple form 
$W(r)=\delta^D(r-r_s)$, because they are independent of the photon propagation 
direction and its path. 
The angular dependence of the integrand is then carried by
the plane wave and this functional dependence can be further
separated by using the partial wave expansion
\beeq
e^{i\bdv{k}\cdot\bdv{x}}=4\pi\sum_{lm}i^l~j_l(kx)~Y_{lm}^*(\Kang)~Y_{lm}(\Vang)~.
\label{eq:wave}
\eneq
The line-of-sight velocity 
$V=v^\alpha e_\alpha$ is independent of the photon path, but depends on
the photon propagation direction; the weight function is 
\beeq
W(r)=\delta^D(r-r_s)(-i\Vang\cdot\Kang)=-\delta^D(r-r_s)\left({1\over k}
{\partial\over\partial r}\right)~,
\eneq
and now it is an operator, acting upon the radial part of
the plane wave. Note that we have explicitly removed the dependence on
the photon propagation direction by making the weight function an operator.
The weight function for the weak lensing convergence~$\kappa$ depends on both
the photon path and its propagation direction, and it is therefore another
operator acting upon the angular part of the plane wave:
\beeq
W(r)=\left({r_s-r\over2~r_s~r}\right)~\ang^2=
-l~(l+1)\left({r_s-r\over2~r_s~r}\right)~,
\eneq
where we have used the relation $\ang^2 Y_{lm}(\Vang)=-l~(l+1)Y_{lm}(\Vang)$.

Finally, for the initial conditions described by a Gaussian
random distribution with $\Delta^2_\phi(k)\propto k^{n_s-1}$,
the auto correlation of large-scale structure can be written as
\beeq
C_l=\langle a_{lm}^*a_{lm}\rangle
=4\pi\int {dk\over k}~\Delta^2_\phi(k)~\II^2(k)~,
\eneq
and the cross correlation of CMB anisotropies with large-scale structure
is 
\beeq
C_l^\times=\langle a_{lm}^\up{cmb*}a_{lm}\rangle 
=4\pi\int{dk\over k}~\Delta^2_\phi(k)
~\TT_l^*(k)~\II(k)~, 
\eneq
where we have defined the angular multipole function of 
large-scale structure in Fourier space
\beeq
\II(k)=\sum_i\int dz\PP(z)
\int_0^{r_s}dr~T_i(k,r)~W_i(r)~j_l(kr)~,
\eneq
and $\TT_l$ is the angular multipole function of CMB anisotropies
(see, e.g., \cite{SEZA96,ZASE97}).

\subsection{Angular correlations}
\label{ssec:ang}
Here we consider a quasar sample without spectroscopic redshift 
measurements used for the cross correlation analysis, such as the
photometric quasar (QSO) sample \cite{HOHIET08}
obtainable from the SDSS. The redshift distribution of the
 sample is assumed to have the standard functional form 
\beeq
P(z)~dz\propto z^\alpha~\exp\left[-\left({z\over z_0}\right)^\beta\right]dz~,
\eneq
with ($\alpha$, $\beta$, $z_0$)=(3, 13, 3.4).
The mean and the peak redshifts of the sample are 2.7 and 3, respectively.

\begin{figure}
\centerline{\psfig{file=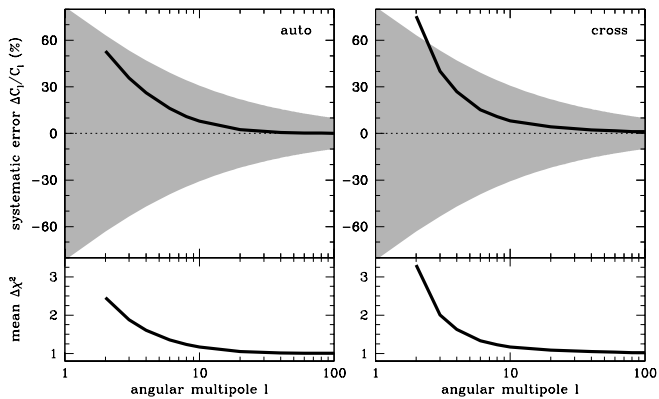, width=3.4in}}
\caption{Systematic errors in theoretical predictions of the
auto correlation ({\it left}) of the QSO sample and its cross
correlation ({\it right}) with CMB temperature anisotropies.
Attached bottom panels show the mean $\Delta\chi^2$ of the measurements,
when only the cosmic variance is considered.
As in the standard practice, the theoretical predictions of the angular
correlations are computed by using 
$\delta_\up{std}=b~\delta_m+(5p-2)\kappa$ with $\delta_m$ in the 
synchronous gauge
and~$\kappa$ in the conformal Newtonian gauge, and the angular correlations 
computed with our full expression for $\dobs$
in Eq.~(\ref{eq:dobs}) are compared to the predictions with $\delta_\up{std}$.
Projection along the line-of-sight suppresses
the large scale modes, where the matter fluctuations in two gauges
in Fig.~\ref{fig:pow} differ substantially. Note that at $l=2$
the correct theoretical prediction is larger by a factor of 1.8
than that one would incorrectly predict in the standard method, and
it is 1.2-$\sigma$ away from the estimated cosmic variance shown as
shaded regions. Since the signal-to-noise ratio of the cross correlation
measurements is largest at low angular multipoles (measurements uncertainties
are large at $l>10$), the systematic errors in the standard method could
bias the inferred cosmology.}
\label{fig:err}
\end{figure}

Figure~\ref{fig:err}
shows the systematic errors in theoretical predictions of the
auto correlation ({\it left}) of the QSO sample and its cross
correlation ({\it right}) with CMB temperature anisotropies, when
the relativistic effects are ignored. Compared to our full expression
in Eq.~(\ref{eq:dobs}), the theoretical predictions in the
standard method are computed by using 
$\delta_\up{std}=b~\delta_m+(5p-2)\kappa$,
where $\delta_m$ is the matter fluctuation in the synchronous gauge
and~$\kappa$ is the convergence in the conformal Newtonian gauge.
We have assumed $b=2$ and $(5p-2)=0.1$ for the QSO sample \cite{HOHIET08},
and the full sky coverage of the survey is assumed for comparison.

In the standard approach to modeling the cross correlation of CMB 
anisotropies with large-scale structure,
the matter fluctuation in large-scale structure
correlates with the integrated Sachs-Wolfe effect 
in CMB anisotropies. 
However, the observed fluctuation field~$\dobs$
in Eq.~(\ref{eq:dobs}) contains numerous new contributions, including
the peculiar velocity, the gravitational potential, and the integrated 
Sachs-Wolfe effect, when written in the conformal Newtonian gauge.
Therefore, when computing the cross correlation with the observed galaxy
fluctuation field, the correlations of the new
contributions are required to be considered in addition to the matter
correlation. For example, the integrated Sachs-Wolfe effect present in both CMB
anisotropies and large-scale structure directly correlates with each
other.

In the conformal Newtonian gauge, $\dz\ll\delta_m$ for the QSO sample,
since the peculiar velocity, the gravitational potential, and the
integrated Sachs-Wolfe effect are of the same order. 
Therefore, when $\dobs$ is computed in the conformal Newtonian gauge,
the correlation of the matter fluctuation~$\delta_m$ contributes most to $C_l$
 and $C_l^\times$ compared to the other numerous contributions,
and the systematic errors in Fig.~\ref{fig:err} arise mainly from the
difference in~$\delta_m$ of the conformal Newtonian and the synchronous
gauges seen in Fig.~\ref{fig:pow}. However, in 
the synchronous gauge, $\dz$ simply results from the integrated Sachs-Wolfe
effect due to the absence of the peculiar velocity and the
gravitational potential, and therefore without accounting for~$\dz$ the
theoretical predictions in the standard method
are underestimated. For the tensor-to-scalar ratio $r=0.1$ at $l=2$,
the tensor contribution is $\sim1\%$ of the matter fluctuation.
We emphasize again that $\dobs$ is gauge-invariant and
it can be computed in any gauges.

As opposed to the dramatic contrast seen in Fig.~\ref{fig:pow} the systematic
errors in the theoretical predictions seem relatively small in 
Fig.~\ref{fig:err}. The main reason is the projection effect in the angular
correlation: each Fourier mode is projected along the line-of-sight and 
the amplitude of $C_l$ is largely
determined by the mode $k\simeq l/r_s$, and slightly larger scale mode
for the cross correlation~$C_l^\times$ due to the cancellation of
two spherical Bessel functions with different distance scales.
This projection effect highly suppresses the largest scale modes
$k\sim1/r_s$ of the sample, reducing the dramatic difference in
the matter fluctuations.
For computing the cross correlation with CMB temperature 
anisotropies, one would in practice need to compute Eq.~(\ref{eq:std})
with $\delta_m$ replaced by the combination $(\delta_m-3~\dz)$
computed in the conformal Newtonian gauge to be consistent with the
calculation of the convergence~$\kappa$.

Finally, we comment on the impact of the systematic errors. At the lowest
angular multipole the accurate theoretical prediction is about a factor
of 1.8 larger than the standard method predicts at the 1.2-$\sigma$
confidence level with the estimated cosmic variance limit. 
Since the cross correlation signals decline rapidly with angular multipole~$l$,
the signal-to-noise ratio of the measurements is determined by the estimated
cosmic variance at $l<10$. The lower theoretical predictions in the standard
method underestimate the cosmic variance, resulting in additional
$\Delta\chi^2$ of a few of the measurements. Considering that the current
detection significance is at the 3-$\sigma$ level for each galaxy sample
\cite{HOHIET08}, these systematic errors could bias the inferred cosmology.

\section{DISCUSSION}
\label{sec:discussion}
We have developed a fully general relativistic description of galaxy clustering
as a cosmological probe --- 
we have derived a covariant expression for the observed galaxy fluctuation
field in a general Friedmann-Lema\^\i tre-Robertson-Walker metric without
fixing a gauge condition and our formalism includes
tensor contributions from primordial gravity waves. The observed volume
element is constructed by using the observed redshift and observed angle in a
homogeneous universe, while the real physical volume element 
given the observables
needs to be constructed by tracing backward the photon geodesic in
an inhomogeneous universe. This discrepancy in the
observable quantities results in a significant modification of the
observed galaxy fluctuation field and provides a key clue for understanding
gauge issues related to the observables.

As our first application,
we have computed the angular auto correlation of the photometric
QSO sample from the SDSS and its cross correlation with CMB anisotropies. 
The cross correlation in the standard method arises from the correlation
of the integrated Sachs-Wolfe effect in CMB anisotropies
and the underlying matter fluctuation of the QSO sample.
However, since there are numerous additional contributions to the observed
QSO fluctuation field, the correlations of the additional terms with
CMB anisotropies need to be considered.
The dominant contribution to the cross correlation
still arises from the matter fluctuation and 
the correct theoretical predictions are larger at
low angular multipoles. The systematic errors in theoretical predictions
are highly suppressed in the angular correlations due to the projection 
effect, but they can result in $\Delta\chi^2$ of a few at low angular
multipoles.

We comment on the possibility to detect primordial gravity waves
using galaxy samples.
Primordial gravity waves affect the observed redshift and position of galaxies
and therefore its effect is imprinted in observed
galaxy fluctuation fields. We find
that the tensor contribution in the cross correlation is 1\%
of the scalar contribution from the matter fluctuation at
the lowest angular multipoles, when the tensor-to-scalar ratio is assumed to
be $r=0.1$ at $l=2$. In general, it is extremely difficult to isolate tensor
contributions in galaxies, because they are completely swamped by scalar
contributions. One possibility is to cross correlate CMB B-mode polarization 
anisotropies with large-scale structure at low angular multipoles as they
should be uncorrelated in the absence of primordial gravity waves.
However, it may not be feasible in practice, as the parity odd quantities
need to be constructed from the observed galaxy samples.

The next application of our formalism is to investigate the effect on the
three-dimensional power spectrum of galaxy samples.
Recently, \citet{DADOET08} showed that the
primordial non-Gaussianity feature can be probed by the scale-dependence
of galaxy bias on large scales and this new development has spurred an
extensive theoretical and observational investigation
\cite{SLHIET08,MAVE08,MCDON08,SELJA09,FISEZA09}.
However, at this large scale, where the
primordial non-Gaussianity feature can be most sensitively probed, 
relativistic effects become substantial and observed quantities
are significantly different from simple theoretical predictions.
Therefore, without proper theoretical modeling of observables,
cosmological interpretation of these measurements would be significantly biased
with the current data, and even more so with galaxy samples from future
dark energy surveys.
Correct modeling of the observed power spectrum would not only
need to account for the discrepancy in the observable
quantities, but also need to account for additional anisotropies arising
from the angle dependence of the observable quantities
(Yoo, Fitzpatrick \& Zaldarriaga in preparation).

\acknowledgments
We acknowledge useful discussions with Daniel Baumann, Antony Lewis,
Jordi Miralda-Escud\'e, Jonathan Pritchard,
and An\v{z}e Slosar.
J.~Y. is supported by the Harvard College Observatory under the
Donald~H. Menzel fund.
A.L.F. is supported by the Department of Energy grant number 
DE-FG02-01ER-40676.
M.~Z. is supported by the David and Lucile Packard, the Alfred~P. Sloan,
and the John~D. and Catherine~T. MacArthur Foundations.
This work was further supported by NSF grant AST~05-06556 and NASA ATP
grant NNG~05GJ40G.

\vfill

\bibliography{ms.bbl}

\end{document}